\documentclass[10pt]{article}
\usepackage{amsmath}
\usepackage{graphicx}
\usepackage{amsfonts}
\usepackage{amssymb}
\usepackage{epsf}
\usepackage{latexsym}

\def\be{\begin{equation}}
\def\ee{\end{equation}}
\def\bea{\begin{eqnarray}}
\def\eea{\end{eqnarray}}
\def\fn{\footnote}
\def\p{\underline{p}}
\def\q{\underline{q}}
\def\r{\underline{r}}

\def\R{\underline{R}}
\def\P{\underline{P}}

\def\N{\underline{N}}
\def\urho{\underline{\omega}}
\def\upi{\underline{\varpi}}
\def\uuA{{\underline{\underline{A}}}}

\def\fa{\mbox{\underline{a}}}
\def\fb{\mbox{\underline{b}}}
\def\aa{\mbox{a}}
\def\bb{\mbox{b}}

\def\ma{\mbox{\scriptsize a}}
\def\dma{\dot{\mbox{\scriptsize a}}}
\def\mb{\mbox{\scriptsize b}}
\def\dmb{\dot{\mbox{\scriptsize b}}}
\def\md{\mbox{\scriptsize d}}
\def\ml{\mbox{\scriptsize l}}
\def\mm{\mbox{\scriptsize m}}
\def\mn{\mbox{\scriptsize n}}
\def\mN{\mbox{\scriptsize N}}
\def\nl{\mbox{l}} 
\def\nm{\mbox{m}} 
\def\nn{\mbox{n}} 
\def\jj{\mbox{j}} 
\def\JJ{\mbox{J}} 
\def\YY{\mbox{Y}}

\def\L{\underline{\cal L}}

\def\Q{\overline{\cal Q}}
\def\st{\mbox{\sffamily T}} 
\def\sq{\mbox{\sffamily Q}} 
\def\sS{\mbox{\sffamily S}}
\def\sG{\mbox{\sffamily G}} 
\def\sv{\mbox{\sffamily V}}
\def\se{\mbox{\sffamily E}} 
\def\un{\mbox{\sffamily E}_{\mbox{\scriptsize Universe}}}

\def\sse{\mbox{\scriptsize{\sffamily E}}} 
\def\ssv{\mbox{\scriptsize{\sffamily T}}} 
\def\sst{\mbox{\scriptsize{\sffamily V}}} 
\def\pa{\partial}
\def\d{\textrm{d}}
\def\cr{\mbox{ } \mbox{\scriptsize{\bf $\times$}} \mbox{ } }
\def\B{\mbox{\tiny B}}

\begin{document}
\begin{titlepage}
\begin{center}

\vspace{.3in}

{\Huge{\bf RELATIONAL PARTICLE MODELS.}}

\vspace{.15in}

{{\bf I. RECONCILIATION WITH STANDARD CLASSICAL AND QUANTUM THEORY}}
\vspace{.3in}

{\Large{\bf Edward Anderson}}

\vspace{.3in}

{\em Department of Physics, P-412 Avadh Bhatia Physics Laboratory,}  

\noindent{\em University of Alberta, Edmonton, Canada, T6G2J1;}   

\noindent{\em Peterhouse, Cambridge, U.K., CB21RD;}

\noindent{\em DAMTP, Centre for Mathemetical Sciences, Wilberforce Road, Cambridge, U.K., CB30WA.}

\end{center}

\vspace{.3in}


\begin{abstract}

This paper concerns the absolute versus relative motion debate.  
The Barbour and Bertotti 1982 work may be viewed as 
an indirectly set up relational formulation of a portion of Newtonian mechanics.    
I consider further direct formulations 
of this and argue that the portion in question -- 
universes with zero total angular momentum, that are conservative and with kinetic terms that 
are (homogeneous) quadratic in their velocities -- 
is capable of accommodating a wide range of classical physics phenomena.  
Furthermore, as I develop in Paper {\bf II}, this relational particle model 
is a useful toy model for canonical general relativity.  

I consider what happens if one quantizes relational rather than absolute mechanics, 
indeed whether the latter is misleading.  
By exploiting Jacobi coordinates, I show how to access many examples of quantized relational 
particle models and then interpret these from a relational perspective.  
By these means, previous suggestions of bad semiclassicality for such models can be eluded. 
I show how small (particle number) universe relational particle model examples display 
eigenspectrum truncation, gaps, energy interlocking and counterbalanced total angular momentum.  
These features mean that these small universe models make interesting toy models for 
some aspects of closed universe quantum cosmology.   
While, these features do not compromise the recovery of reality as 
regards the practicalities of experimentation in a large universe such as our own.  

\end{abstract}


\vspace{.3in}



\end{titlepage}

\section{Introduction}

Newton's formulation and conceptual framework for mechanics is based on his notions of 
absolute space and absolute time \cite{Newton}.    
However, Leibniz \cite{L}, Berkeley \cite{B} and Mach \cite{M} envisaged that mechanics should 
rather be {\sl relational} i.e feature solely relative distances, angles and times.  
Despite these occasional criticisms, no concrete realizations of such mechanical theories were 
achieved, and absolutism took hold.   
So while Lagrange \cite{Lag} and Jacobi \cite{Jac} invented notable techniques for the study of 
$N$ body Newtonian mechanics (NM): working in the barycentre frame, using relative particle separation coordinates, 
using the relative particle cluster separation Jacobi coordinates that diagonalize the relative formulation's kinetic term, 
eliminating one angular momentum and using the Jacobi form for the action, they appear not to have assembled 
these techniques into a relationalist program \cite{Relper}.  
The first elements of that appear in Dziobek \cite{Dziobek} and Poincar\'{e} \cite{Poincare}. 
A concrete relational synthesis was attained by Barbour and Bertotti in 1982 (BB82) \cite{BB82} 
which I present in Sec 2. 
This has also been studied in \cite{BS, Rovelli, Smolin, B94I, B94II, Buckets, LB, EOT, Gergely, 
GergelyMcKain, RWR}, reviewed in \cite{Kuchar92, Kiefer, Landerson} and philosophized about in 
\cite{Barbourphil, Buckets, EOT, Pooley}.  
It consists of a classical relational reformulation of a {\sl portion of NM}.  
BB82-type formulations may be perceived to have an undesirable indirectness in that 
while their derivation ends up free of absolute space, it proceeds indirectly via what 
may be viewed as absolutist notions (even if BB82 select these purely on grounds of 
convenience among various ways of conceptualizing in terms of configuration spaces).    
It is worth commenting that while it is not logically necessary\fn{Starting 
on relational premises, one can also arrive at {\sl distinct} theories of particle mechanics, 
see Sec 3.} 
for a relational viewpoint to lead\fn{Such 
a recovery of NM from different premises parallels Wheeler's promotion of many routes to 
general relativity (GR) rather than just Einstein's.  
See Paper {\bf II} of this series \cite{PaperII} for references and further discussion.}   
to a recovery of NM, it is nevertheless interesting for a resolution of the absolute versus 
relative motion debate to have such a feature.

In this paper I examine and improve three key features \cite{Panderson} of BB82 relational particle mechanics (RPM).  
Firstly, it {\sl can} also be cast in a {\sl directly relational form}, 
which I attain in Jacobi action relative particle separation coordinate form in Sec 3\fn{See also 
Lynden-Bell \cite{LB} and Gergely \cite{Gergely} for earlier formulations.} 
and furthermore recast in terms of {\sl relative Jacobi coordinates} 
in Sec 4.  
Jacobi coordinates are particularly well-suited to RPM and are mathematically-central to this paper.      

Secondly, by a `portion' I mean the restriction to closed systems/universes 
of zero total angular momentum (AM) $\L = 0$, 
that are conservative and have a kinetic term that is homogeneous quadratic in the velocities.    
It is therefore important to investigate whether these are major restrictions.  
In Sec 5 I argue that this portion accommodates much, by tricks which  
counterbalance the {\sl sub}system of interest with other subsystems elsewhere.   
This portion also admits a {\sl non}homogeneous quadratic extension.  

Thirdly, I begin to investigate the quantum implications of shifting from 
an absolutist to a relationalist viewpoint.  
This is particularly relevant given how GR and QM incorporate incompatible notions of time.  
This `Problem of Time' and its conceptual ramifications are good reasons (see e.g. 
\cite{Battelle, Kuchar80, Kuchar92, Isham93, KucharPOTother, B94II, EOT})  
why these two areas of theoretical physics have not been successfully combined 
insofar as we do not have a satisfactory account of quantum GR 
(nor a fully satisfactory account of any distinct theory of quantum gravity at all).  
In particular, the traditional framework of QM is based on 
Newton's formulation of particle mechanics, absolute space and time and all, 
while GR, as well as being a successful reconciliation of SR and Newtonian gravity 
and a successful improvement of the theory of gravitation, 
is also Einstein's attempt to free physics of absolute structure.  
Many theoretical physicists consider this feature of GR to be 
an important lesson about how the universe works, to the extent that   
quantum gravity schemes that, contrarily, have extraneous `background dependence' 
\cite{bigcite, Isham93, Kiefer, Dirac}\fn{One 
should also take care to note that background (in)dependence can take on a different meaning 
in the string- and M-theory literature, namely vacuum choice independence of string perturbations.} 
can be regarded as not yet {\sl conceptually} satisfactory.  
Indeed one major goal of nonperturbative 
M-theory is to attain this background independence.  

Now, given that the relational reformulation of $\L = 0$ NM  
{\sl is itself a successful abolition of absolute structure}, 
the following interesting question arises.
Q1: Does this give a different sort of QM from that of the traditional absolutist approach?  
A further reason for this study is that the BB82 RPM approach to 
$\L = 0$ NM closely parallels \cite{BB82, B94I, RWR, EOT, Landerson} 
that to the geometrodynamical formulation of spatially compact without boundary GR 
\cite{BSW, RWR, Vanderson, TP}.  
Indeed, \cite{BS, Kuchar92, B94I, B94II, EOT, Kiefer} have considered BB82 RPM 
as a useful toy model arena\fn{\cite{SIPP, ABFKO} 
and {\bf II} also view a more recent scale invariant model as useful in this respect.}    
for the investigation of quantum gravity and quantum cosmology issues such as 
the Problem of Time and difficulties with closed universes.  
Note however that there is a perceived obstruction to this program.
While Barbour and others have been advocating BB82 RPM 
as a resolution of the absolute versus relative motion debate for some years, the 
Barbour--Smolin (BS) preprint \cite{BS} {\sl also} claims that quantizing BB82 does 
not give a good semiclassical limit on two counts. 

\noindent BS1: that small masses affect the QM spread of large masses.     

\noindent BS2: Due to constituent subsystems being interlocked by additional restrictions.

\noindent This position is followed up by arguing against the applicability 
of standard quantization procedures, leading them instead to `seek a radical alternative'.  
\cite{B94II, EOT} may be seen as the start of Barbour's search for this -- a particular 
{\sl consistent records} program.  

Nevertheless, in this paper I begin by addressing Q1.  
This subdivides into whether different formal mathematics becomes involved and whether there are 
interpretational differences.  
I principally consider simple 1-$d$ models, 
regarding extension to $d > 1$ models where possible as a useful bonus.  
I do so by use of relative Jacobi coordinates,\fn{While this paper provides 
relationalist motivation underlying this technique, N.B.  
that the technique itself is well-established in the molecular physics 
literature \cite{LJR} due to its practical usefulness.} 
noting that these map many relational quantum problems to problems whose formulation 
is known from the standard quantization of the absolutist formulation of NM.  
Thus the first steps 
of the study of these simplest RPM models 
do {\sl not} involve formal mathematical differences, permitting one to benefit from 
separation and special function techniques.  
Thereby I can cover a wide range of RPM models 
for the minimally relationally-nontrivial case of 3 particles.   
These go well beyond BS's (piecewise) constant potentials in 1-d to additionally solve 
a number of interparticle harmonic oscillator and Newton--Coulomb potential 
problems including those presented in Sec 7.   

The first salient difference is interpretational.  
Spreads are now in terms of relational quantities (relative separations suffice in 1-d), 
from which perspective I show that objection BS1 is misplaced.  
The second is mathematical, but turns out not to spoil the above standard techniques by being treatable 
{\sl after} deploying these techniques.  
This involves the collective of subsystems having interlocked energy and counterbalancing AM, 
and the subsystem eigenspectra sometimes exhibits truncations and gaps.  
These features are a further development of the mathematical observations which led to objection BS2, 
but I also develop further what the implications of these are.  
For universes that have a large particle number and a diverse content 
(such as free particles and of both positive- and negative-potential subsystems), 
gaps need no longer occur, truncations become acceptable, while   
interlocking and counterbalancing are obscured by the practicalities of 
experimentation involving only a small subsystem.  
As our own universe 
is large and diverse in content, the {\sl `recovery of reality'} is thereby not compromised by BS2.   
This opens the way for semiclassical exploration of these models, which I consider further, 
alongside the aforementioned consistent records formulation, in {\bf II}.7--8.
On the other hand, {\sl small universe models} do exhibit these  
interesting theoretical effects whereby they are useful toy models for understanding 
subtle ramifications that arise from considering {\sl closed}-universe quantum cosmologies.

\section{Barbour--Bertotti 1982 formulation in terms of auxiliary variables}

Let $\sq$ be the na\"{\i}ve $Nd$-dimensional configuration space in $d$ 
dimensions for $N$ particles whose positions are $q_{\alpha A}$.\fn{Lower-case Greek letters 
running from $1$ to $d$ are 
spatial indices (sometimes I use the underline notation instead of the index notation for these); 
I shall develop models for $d = 1$ to $3$.
Upper-case Latin letters running from $1$ to $N$ are particle labels, and 
lower-case Latin letters running from $1$ to $N - 1$ label relative particle (cluster) separations.}  
Jacobi actions are used, which implement temporal relationalism since they are 
{\sl reparametrization-invariant} in the label-time $\lambda$.  
The actions considered in this article are for particle models 
whose kinetic term $\st$ which is homogeneous quadratic in its constituent velocities. 
These take the form\fn{See Sec 4 for a generalization of this form required for the 
inclusion of physics with linear kinetic terms.} 
\be
\sS_{\mbox{\scriptsize{Jacobi}}} = 2\int\d\lambda \sqrt{    (  \se- \sv  )\st      } \mbox{ } , 
\label{Jac}
\ee
where $\sv$ is the potential term and $\se$ is the total energy, taken to be a fixed constant 
`energy of the universe' $\un$.  
Note that the more usual particle mechanics action,  
$\sS_{\mbox{\scriptsize{Euler--Lagrange}}} = \int\d t (\st- \sv)$, 
can indeed be cast \cite{Lanczos} into the incipient Jacobi form (\ref{Jac})  
by adjoining the Newtonian time $t$ to $\sq$, then noting that $\dot{t} = \d t/\d\lambda$ alone features in the action and 
subsequently eliminating it by Routhian reduction.  

The classical particle mechanics notion of spatial relationalism is implemented by passing to 
a suitable notion of arbitrary frame.\fn{This account is a further development 
of the matter rather than following the formulation in the BB82 paper verbatim.}  This is achieved 
by the introduction of a translational auxiliary $d$-vector $\aa_{\alpha}(\lambda)$ and whichever 
rotational auxiliary corresponds to $d$: none for $d = 1$, a scalar b$(\lambda)$ for $d = 2$ or a 
3-vector b$_{\alpha}(\lambda)$ for $d = 3$, so that the $A$th particle's position $q_{\alpha A}$ is 
replaced by 
\be
\&_{\ma,\mb}q_{\alpha A} \equiv 
q_{\alpha A} - \aa_{\alpha} -  {\epsilon_{\alpha}}^{\beta\gamma}\bb_{\beta} q_{\gamma A} 
\mbox{ } .  
\label{abcorr}
\ee
Whereas the 1- and 2-$d$ cases can be written in intrinsically 1- and 2-$d$ notation, 
I present all 3 cases together by use the extraneous 3-$d$ $\epsilon$ symbol,  
under the provisos that 
\be
\bb_{\alpha} = 0 \mbox{ for } d = 1 
\mbox{ } \mbox{ and } \mbox{ }
\bb_{\alpha} = (0,0,\bb_3) 
\mbox{ } , \mbox{ } 
\bb \equiv \bb_3 
\mbox{ } , \mbox{ } 
{\cal L}_{\alpha} = (0, 0, {\cal L}_3) 
\mbox{ } , \mbox{ } 
{\cal L} \equiv {\cal L}_3  \mbox{ for } d = 2.
\label{provisos}
\ee
While this may look like {\sl doubling} the allusions to absolute space, 
$\sq \longrightarrow \sq \times \mbox{Eucl($d$)}$ [for Eucl($d$) the Euclidean 
group of translations and whichever rotations exist in dimension $d$], 
I reassure the reader that this doubling of redundancy leads promptly below to 
the removal of the redundancy in the fashion familiar in gauge theory.   
I proceed by requiring the action to be built as best as possible out of objects 
that transform well under $\lambda$-dependent Eucl($d$).  
In particular, defining the {\sl na\"{\i}ve} or {\sl Lagrange relative coordinates} (relative separations 
between particles)   
$r_{\alpha AB} \equiv q_{\alpha A} - q_{\alpha B}$, 
\be
\sv = \sv(|r_{\alpha AB}| \mbox{ alone}) \mbox{ } , 
\label{Vrel}
\ee
wherein the auxiliary corrections straightforwardly cancel each other out.  
The situation with the kinetic term is more complicated. 
$\frac{\pa}{\pa\lambda}$ is not a tensorial operation under $\lambda$-dependent Eucl(d) transformations.  
As explained in \cite{Stewart}, $\frac{\pa}{\pa\lambda}$ should rather be seen as the Lie derivative $\pounds_{\frac{\pa}{\pa\lambda}}$ 
in a particular frame, which transforms to the Lie derivative with respect to `$\frac{\pa}{\pa\lambda}$ corrected additively by 
generators of translations and rotations.'  
This gives a kinetic term of the form 
\be
\st(\q_I, \dot{\q}_J, \dot{\fa}, \dot{\fb}) = \frac{1}{2}\sum_{A = 1}^{N}m_A\delta^{\alpha\beta} 
\left(
\&_{\dma, \dmb}{q}_{\alpha A}
\right)
\left(
\&_{\dma, \dmb}{q}_{\beta A}
\right) \mbox{ } .
\label{T}
\ee
So, finally, the proposed action is 
$\sS_{\mbox{\scriptsize{Jacobi}}}(\q_I, \dot{\q}_J, \dot{\fa}, \dot{\fb})$  
of form (\ref{Jac}) with (\ref{Vrel}) and (\ref{T}) substituted into it.

The momenta conjugate to the $q_{\alpha A}$ are 
\be
p^{\alpha A} = \sqrt{\frac{\se - \sv}{\st}}m_A\delta^{\alpha\beta}\&_{\dma, \dmb}\dot{q}_{\beta A} 
\mbox{ } .  
\ee
By virtue of the particular reparametrization invariance of the Lagrangian, these obey the primary quadratic constraint
\be
{\cal Q} \equiv \sum_{A = 1}^{N}\frac{1}{2m_{A}}\delta_{\alpha\beta}p^{\alpha A}p^{\beta A} + \sv = \se \mbox{ } .
\ee
The secondary linear constraints, 
\be
{\cal P}^{\alpha} \equiv \sum_{A = 1}^{N}p^{\alpha A} = 0 
\mbox{ } \mbox{ } \mbox{ (zero total momentum constraint) } 
\label{ZM}
\ee
and whichever portion of 
\be
{\cal L}^{\alpha} \equiv \sum_{A = 1}^{N} {\epsilon^{\alpha\beta}}_{\gamma} q_{\beta A}  p^{\gamma A} = 0 
\mbox{ } \mbox{ } \mbox{ (zero total AM constraint) }
\label{ZAM}
\ee
is relevant in the corresponding dimension,  
follow from variation of the cyclic auxiliary coordinates $\mbox{a}_{\alpha}$ and $\mbox{b}_{\alpha}$.  
These constraints obey the Poisson bracket algebra  
\be
\{{\cal P}^{\alpha}, {\cal P}^{\beta}\} = 0 
\mbox{ } , \mbox{ } 
\{{\cal P}^{\alpha}, \Q \} = 0
\mbox{ } , \mbox{ }
\{\Q, \Q\} = 0 \mbox{ } , \mbox{ }
\{{\cal P}^{\alpha}, {\cal L}^{\beta} \} = {\epsilon^{\alpha\beta}}_{\gamma}{\cal P}^{\gamma} 
\mbox{ } , \mbox{ }
\{{\cal L}^{\alpha}, {\cal L}^{\beta}\} = {\epsilon^{\alpha\beta}}_{\gamma}{\cal L}_{\gamma} 
\mbox{ } , \mbox{ }
\{{\cal L}^{\alpha}, \Q\} = 0
\mbox{ } 
\label{first6}
\ee
for $\Q = {\cal Q} - \se$.  
As this is closed, there are no further constraints.  
The constraints (\ref{ZM}, \ref{ZAM}) then signify 
that the physics is not on the doubly redundant 
configuration space $\sq \times \mbox{Eucl($d$)}$ but on the quotient space $\sq/\mbox{Eucl($d$)}$, 
thus indeed rendering absolute space irrelevant.

It is worth counting to establish which are the minimal nontrivial dynamics examples.  
There are $dN$ degrees of freedom in the $d$-dimensional NM of $N$ particles.  
There are $d$ translations and $d(d - 1)/2$ rotations which make up $d(d + 1)/2$ irrelevant motions.  
So the $d$-dimensional $N$-particle BB82 RPM has $d(2N - d - 1)/2$ degrees of freedom.
Furthermore, one wants to express one change in terms of another change rather than in terms of an 
arbitrarily-reparametrizable label-time.  
Thus dynamical nontriviality dictates $N$ and $d$ be such that 
$d(2N - d - 1)/2 > 1$ which gives $N \geq 3$ both in 3-$d$ and in 1-$d$.  
In this article I focus principally on the simplest nontrivial dynamics: $N = 3$.

\section{Passage to direct formulation in terms of relative variables alone}  

In contrast to the above formulations, there is a distinct, non-Newtonian, failed and often 
rediscovered theory\fn{Sometimes called BB 1977 theory \cite{BB77}, as described in \cite{Buckets}, 
this theory is first known to have been discovered by Hoffmann, and was rediscovered by Reissner 
and then by Schr\"{o}dinger.  While this theory benefits from being able to explain the anomalous 
perihelion shift of Mercury (while still being a nonrelativistic particle theory !), it 
additionally predicts an level of mass anisotropy that is unacceptable given the results of the 
Hughes--Drever experiment \cite{HughesDrever}.  I note furthermore that there are yet other such 
RPM's \cite{SIPP, LB}, but these either disagree with observation, or have not yet been 
studied enough to know whether they disagree.} formulated 
directly in terms of the $r_{IJ}$ and without auxiliaries referring to absolute space.  
I should point out that this theory has a kinetic term that is exceptionally simple,  
\be
\st = \frac{\delta^{\alpha\beta}\delta^{IK}\delta^{JL}}{|r_{PQ}|}\dot{r}_{\alpha IJ}\dot{r}_{\beta IK} 
\mbox{ } ,
\label{BB77T}
\ee  
which is probably one reason why this theory has been rediscovered so many times.  
The question then arises whether BB82 RPM can {\sl also} be cast as a presumably more complicated 
direct formulation of this kind, whereupon not even indirect or unphysical reference to 
absolute space is required.  
I address this below, partly foreshadowed by work of Lynden-Bell  
\cite{LB}, Gergely \cite{Gergely} and Jacobi.  

I use that the $d = 3$ working contains everything under the provisos (\ref{provisos}).   
and then comment on each individual case $d = 1, 2, 3$ as these differ significantly.    
I begin with $\sS_{\mbox{\scriptsize Jacobi}}(\q_I, \dot{\q}_J, \dot{\fa}, \dot{\fb})$.  
I define $\tilde{m}$ as the total mass ${\sum_{I = 1}^{N} m_I}$.     
I eliminate $\dot{\aa}_{\alpha}$ from its own variational equation  
\be
\dot{\aa}_{\alpha} = \frac{1}{\tilde{m}}\sum_I m_I
\left(
\dot{q}_{\alpha I} - {\epsilon_{\alpha}}^{\beta\gamma}\dot{\bb}_{\beta}q_{\gamma I} 
\right)
\ee 
[the Lagrangian counterpart of the Hamiltonian expression (\ref{ZAM})]. 
This results in the (semi-)eliminated action 
\be
\sS^*(\r_{IJ}, \dot{\r}_{KL}, \dot{\fb}) = 
2 \int \d \lambda\sqrt{(\se - \sv(|\r_{GH}|)\st^*(\dot{\r}_{IJ}, \mbox{ } \dot{\fb} \cr \r_{KL})}
\mbox{ }
\mbox{for} 
\mbox{ }
\st^*(\dot{\r}_{KL}, \mbox{ } \dot{\fb} \cr \r_{IJ}) = 
\sum_{I \mbox{ } <}\sum_J \frac{m_Im_J}{2\tilde{m}}|\dot{\r}_{IJ} - \dot{\fb} \cr {\r}_{IJ}|^2.
\label{semi}
\ee
I next eliminate $\dot{\bb}_{\alpha}$ from its own variational equation\fn{This gives the same 
result as if $\dot{\fb}$ is eliminated first, or indeed as if 
$\dot{\fa}$ and $\dot{\fb}$ are eliminated simultaneously.}   
\be
\dot{\bb}^{\alpha} = (\stackrel{\B}{I}{}^{-1}(\r_{IJ}))^{\alpha\beta}
                      \stackrel{\B}{L}_{\beta}(\r_{IJ}, \dot{\r}_{KL})   
\label{bvar}
\ee
where $\stackrel{\B}{I}_{\alpha\beta}$, $\stackrel{\B}{L}_{\alpha}$ are the barycentric  
inertia tensor and AM respectively, which are easily castable into relative separation coordinate form:
\be
\stackrel{\B}{I}_{\alpha\beta}(\r_{IJ}) = \sum_{I \mbox{ } <}\sum_J \frac{m_Im_J}{\tilde{m}}
\left(
|\r_{IJ}|^2\delta_{\alpha\beta} - \r_{\alpha IJ}{\r}_{\beta IJ}
\right)  
\mbox{ } , \mbox{ } 
\stackrel{\B}{L}_{\alpha}(\r_{IJ}, \dot{\r}_{KL}) = {\epsilon_{\alpha}}^{\beta\gamma}
\sum_{I \mbox{ } <}\sum_J \frac{m_Im_J}{\tilde{m}}\r_{\beta IJ} \dot{\r}_{\gamma IJ} 
\mbox{ } . 
\label{relf}
\ee 
This results in the eliminated action 
\be
\sS^{**}(\r_{PQ}, \dot{\r}_{ST}) =
2\int \d \lambda \sqrt{        (\se - \sv(|\r_{NO}|))\st^{**}(\r_{PQ}, \dot{\r}_{ST})           } 
\label{bmthgg}
\ee
for
\be
\st^{**}(\r_{PQ}, \dot{\r}_{ST})  = \st^{*}(\dot\r_{PQ}, 0)  
 - \frac{1}{2}\stackrel{\B}{L}_{\alpha}(\stackrel{\B}{I}{}^{-1})^{\alpha\beta}\stackrel{\B}{L}_{\beta} 
\mbox{ } . 
\label{Trel}
\ee

(\ref{bmthgg}) is the formulation of the stated portion of NM 
cast entirely in relative terms for $d = 3$.  
For $d = 1$, $\fb = 0$ so the working stops at (\ref{semi}).  
For $d = 2$, (\ref{bvar}) simplifies considerably since $\stackrel{\B}{I}_{\alpha\beta}$ 
becomes a mere scalar 
\be
\stackrel{\B}{I}(\r_{IJ}) \equiv \stackrel{\B}{I}_{33}(\r_{IJ}) 
\stackrel{2-d}{=} \sum_{I \mbox{ } \mbox{ } <}\sum_{J}\frac{m_Im_J}{\tilde{m}}|{\r}_{IJ}|^2
\mbox{ } ,
\ee 
so one  has trivial explicit invertibility and the second term in (\ref{Trel}) simplifies 
considerably [see also (\ref{Jenga})].

\section{Coordinate improvements: Jacobi coordinates}

As written, my expressions depend on all of the $\r_{IJ}$'s rather than on an $(N - 1)$-member basis. 
I first redress this by noting that 
\be
\r_{IJ}, I = 1 \mbox{ to } N - 1, J = I + 1 \mbox { form a basis for the relative separation coordinates, }
\label{basis}
\ee
and recasting the previous section's symmetric but redundant `Lagrangian' expressions asymmetrically 
yet nonredundantly in terms of these alone [e.g using $\r_{13} = \r_{12} + \r_{13}$].   

Next I consider mapping the $N$ $\q_I$ position vector coordinates linearly to a maximal 
number $N - 1$ vector relative particle (cluster) separation  
coordinates $\urho_i$ and 1 vector absolute coordinate 
$\urho_{\mN}$.  The relative coordinates therein can be recognized since then from (\ref{basis}) 
\be
\urho_I \mbox{ for a given component $I$ is relative iff } 
\urho_I = \sum_{J= 1}^{N}\uuA_{IJ}\q_J \mbox{ for } \uuA_{IJ} \mbox{ constants such that }
\sum_{J=1}^N\uuA_{IJ} = 0 
\mbox{ } .
\label{test}
\ee
If one has any such set of coordinates containing a maximal number of relative coordinates 
$\urho_i$, $i = 1$ to $N - 1$, the final coordinates  
$\urho_{\mN}$ being no true relative coordinates, the zero total momentum constraint of Sec 2 gives 
$$
0 = \sum_{I = 1}^{N}\p^I 
  = \sum_{I = 1}^{N}\sum_{J = 1}^{N} \frac{\pa \urho_J}{\pa \q_I} \upi^I  
  = \sum_{I = 1}^{N}\sum_{J = 1}^{N} \uuA_{IJ}\upi^I 
  = 0 + \sum_{I = 1}^{N}\uuA_{\mN I}\upi^{\mN} \Rightarrow 
$$
\be
\upi^{\mN} = 0 \mbox{ } , 
\label{irrel}
\ee
where $\upi^I$ are the momenta corresponding to the new set of coordinates, and using (\ref{test}) in 
the last 2 steps.  Thus the zero momentum constraint can be absorbed by passing 
to any such coordinate system, and this causes the absolute coordinates $\urho_{\mN}$ therein 
to drop out of all remaining equations, which thereby involve relative coordinates alone.       

Furthermore, I wish $\st^*$ not only to be in terms of a basis of relative coordinates 
but also to be diagonal.  
This will be particularly useful in this paper when it comes to quantization, and is true of the 
{\sl Jacobi coordinates} $\R_I$ \cite{LJR, Marchal}, which exist for all relevant $N$ (i.e $N \geq 3$).    
As among the ($N$ arbitrary) Jacobi coordinates, without loss of generality $\R_i$ $i = 1$ to $N - 1$ 
are relative, I set these to be the particular relative coordinates I use.  
Because of the existence of the zero momentum constraint, which may be cast in the form (\ref{irrel}), 
it is irrelevant which single absolute coordinate vector I adjoin to these as the constraint will in any case
wipe out this choice, so I can choose my last coordinate to be $\q_{\mN}$ rather 
than $\R_{\mN}$ for simplicity of computation.  
I then find that the relative Jacobi coordinates not only diagonalize $\st$ but 
have a large number of further properties which are particularly well-suited to the study 
of RPM.  
Namely, the form-preservation under the $\q_I$ {\it space to} $\R_i$ {\it space map} of $\st$ 
(both as a function of velocities and of momenta), the moment of inertia $J$,  
$\stackrel{\B}{I}_{\alpha\beta}$, $\L_{\alpha}$ and $\stackrel{\B}{L}_{\alpha}$;   
see also Sec {\bf II}.4 for one more such preserved object and Secs 6 and {\bf II}.6 for major  
applications of these properties.

As a first application of relative Jacobi coordinates, I recast the indirect formulation 
of Sec 2 as the action 
\be
\sS^{*}(\R, \dot{\R}, \dot{\fb}) = 2\int\d\lambda \sqrt{    (  \se- \sv(|\R_{i}|, \R_j\cdot \R_k)
                                  \st^{*}(\dot{\R_i}, \mbox{ } \dot{\fb} \cr \R_k)    } 
\mbox{ } 
\mbox{ for } 
\mbox{ }
\st^*(\dot{\R_i}, \dot{\fb} \cr \R_k) = 
\sum_{i = 1}^{N - 1}\frac{M_i}{2}|\dot{\R}_i - \dot{\fb} \cr \R_i|^2 
\mbox{ } .  
\ee
With $\underline{{\cal P}} = 0$ reducing to the absense of any additional absolute coordinates, 
the surviving constraints are\fn{Here, the $M_i$ are redefined masses in terms of the 
original masses, $m_J$.  I do not provide $M_i = M_i(m_J)$ relations here as these in any case 
depend on normalization convention, and Jacobi coordinates are nonunique for $N > 3$.}  
\be
\Q = \sum_{i = 1}^{N - 1}\frac{\P_i^2}{2M_i} + \sv - \se
\label{HamT}
\ee
and 
\be
{\cal L}_{\alpha}(\R_i, \dot{\R}_j) = 
{\epsilon_{\alpha}}^{\beta\gamma}\sum_{i = 1}^{N - 1}R_{\beta i} P^i_{\gamma} 
= \sum_{i = 1}^{N - 1}L_{\alpha i} = 0 \mbox{ } . 
\label{DamT}
\ee 
Note how employing relative Jacobi coordinates retains the separability of the zero AM constraint 
while preserving the form of the individual AM operators.  

As a second application, I recast the direct formulation of Sec 3 dimension by dimension. 
For $d = 1$, 
\be
\sS^{**}[R_i] = 2\int \d \lambda \sqrt{    (  \se - \sv(R_{i}) 
                                  \st^{**}(\dot{R_i})    } 
\mbox{ } \mbox{ for } \mbox{ } 
\st^{**}(\dot{R_j}) = \sum_{i = 1}^{N - 1} \frac{M_i}{2}\dot{R}_i^2 \mbox{ } .
\ee
So these coordinates are diagonal and also  $d = 1$ ensures that this formulation is not just 
relative but also fully relational, because $\sS = \sS[|\R_{i}|]$ {\sl alone}.  

For $d = 2$, 
\be
\sS^{**}[\R_{i}] = 2\int \d \lambda \sqrt{    (  \se - \sv(|\R_{i}|, \R_j\cdot \R_k ) 
                                            \st^{**}(\R_i, \dot{\R_j})        }  
\ee
for 
\be 
{\st}(\R_i, \dot{\R_j}) = 
\sum_{i = 1}^{N - 1} \frac{M_i}{2}|\dot{\R}_i|^2 - 
\sum_{i = 1}^{N - 1}\sum_{j = 1}^{N - 1}
\frac{M_iM_j}{2\stackrel{\B}{I}}
\epsilon^{\alpha\beta}\epsilon^{\gamma\delta}R_{i\alpha}\dot{R}_{i\beta}R_{j\gamma}\dot{R}_{j\delta}
\mbox{ } \mbox{ and } \mbox{ } 
\stackrel{\B}{I} = \sum_{i = 1}^{N - 1}{M_i}|\R_i|^2
\mbox{ } .
\label{Braz}
\ee
But 
\be
\epsilon^{\alpha\beta}\epsilon^{\gamma\delta}R_{i\alpha}\dot{R}_{i\beta}R_{j\gamma}\dot{R}_{j\delta} 
= (\R_1 \cdot \R_2)(\dot{\R}_1 \cdot \dot{\R}_2) - (\R_1 \cdot \dot{\R}_2)(\R_2 \cdot \dot{\R}_1) 
\mbox{ } , 
\ee 
which is made out of relative angles\fn{$\left\{   |\R_i|^2, 
\mbox{ } 
\mbox{arccos}\left(\frac{\R_j \cdot \R_k}{|\R_j||\R_k|}\right)   
\right\}$ 
and 
$\left\{   |\R_i|^2, 
\mbox{ } 
\R_j \cdot \R_k  
\right\}$ clearly span the same set of functions, which are the truly relational quantities.} 
and relative separations, and so the Jacobi coordinates serve to give a fully relational formulation.  
Alternatively [to compare it with (\ref{BB77T}), with \cite{Gergely} and for future reference]
(\ref{Braz}) can be written in terms of an (inverse) configuration space metric 
\be 
\st(\R_{i}, \dot{\R}_{j})  =  \frac{1}{2} 
\sum_{i = 1}^{N - 1}\sum_{j = 1}^{N - 1}
(\sG^{-1})^{\alpha\beta ij}\dot{R}_{\alpha i}\dot{R}_{\beta j} 
\mbox{ } , \mbox{ }
(\sG^{-1})^{\alpha\beta ij} = M_i\delta^{\alpha\beta}\delta^{ij} - 
\frac{    M_iM_j    }{      \sum_{k = 1}^{N - 1} M_k|\R_k|^2    }
\epsilon^{\alpha\beta}\epsilon^{\gamma\delta}R_{\gamma}^{i}R_{\delta}^{j}  
\mbox{ } .
\label{Jenga}
\ee
[For $d = 1$ the metric is flat].  

For $d = 3$, 
\be
\st^{**}(\R_i, \dot{\R_j}) = \st^{*}(\dot{\R_i}, 0) - \frac{1}{2}
\stackrel{\B}{L}_{\alpha}
(\stackrel{\B}{I}{}^{-1})^{\alpha\beta}
\stackrel{\B}{L}_{\beta}
\mbox{ } , 
\ee
\be
\stackrel{\B}{I}_{\alpha\beta} = \sum_{i = 1}^{N - 1}M_i
\left(
|\R_i|^2\delta_{\alpha\beta} - R_{\alpha i}R_{\beta i}  
\right)
\mbox{ } \mbox{ and } \mbox{ } 
\stackrel{\B}{L}_{\alpha}(\R_i, \dot{\R}_j) = 
{\epsilon_{\alpha}}^{\beta\gamma}\sum_{i = 1}^{N - 1}M_iR_{\beta i}\dot{R}_{\gamma i}\mbox{ } . 
\ee
The inverse configuration space metric is now
\be
(\sG^{-1})^{\alpha\beta ij} = 
M_i\delta^{\alpha\beta}\delta^{ij} -     M_iM_j     
{\epsilon_{\gamma}}^{\delta\alpha}(\stackrel{\B}{I}{}^{-1})^{\gamma\mu}{\epsilon_{\mu}}^{\lambda\delta} 
R_{\delta}^i R_{\lambda}^j \mbox{ } .  
\label{Janga}
\ee
Note the complication through the presence of the nontrivial inverse of the inertia tensor $I^{-1}$.    
I currently can do no better than express this as relational variables plus angles between 
Jacobi vectors and principal directions of the system's inertia quadric. 
This underscores the approach in practice for papers {\bf I} and {\bf II}: work on techniques for 
the 1-$d$ problems since these will turn out to already exhibit many of the interesting features of 
RPM models; direct extensions of these methods to $d > 1$ are furthermore worth considering even 
though they will sometimes be a semi-relational `halfway house' rather than fully relational methods.  
Moreover, some of the remaining interesting features appear in the scale-invariant 1-$d$ set-up 
(see paper {\bf II}).    

The above are the last formulations of the 1-$d$, 2-$d$ and 3-$d$ $\L = 0$ conservative homogeneous 
quadratic portion of NM in this paper.  
[Sec {\bf II}.4 contains yet further formulations and 
study of the configuration space metrics (\ref{Jenga}, \ref{Janga})].
I next address the question of whether focussing on this portion of NM is a significant restriction.

\section{How restrictive is considering only this portion, at a classical level?}

BB82 RPM and $\L = 0$ NM give coincident physical predictions. 
If $\L \neq 0$ momentum physics is under study, one can always represent it as a {\sl subsystem} 
of a $\L = 0$ universe \cite{BB82, LB}, so that the restriction to $\L = 0$ universes is by no means 
as severe as might be na\"{\i}vely suggested.  

One concern is that the viability of this rests on the $N$-body problem's theoretical framework 
being such that initially distant subsystems do not fall together below any desired finite timescale 
internal to the subsystem under study.  
Reversely, such falling together requires the unboundedness of some velocities and hence 
of $\st$ and hence of $\sv$.      
This is termed a {\sl singularity}.  
While it used to be thought that this could only occur in situations involving collisions,  
it has been shown that for 5 bodies it is possible to attain this by merely coming 
arbitrarily close to collisions \cite{Xia}.  
It is still mere conjecture however that singular solutions are of measure zero in the 
set of all solutions.  
In any case, sufficiently accurate physical modelling of the universe acts to prevent arbitrarily 
distant subsystems from falling together in finite time: astrophysics with realistic matter will 
experience short-range forces interfering with the potential being arbitrarily negative or GR (which 
respects a positive energy theorem) will take over, while SR will bound infall velocities.  

Can one test whether $\L = 0$ or $\neq 0$ in our universe?  
It is true that $\L = 0$ and $\neq 0$ (sub)systems are capable of evolving qualitatively differently 
as exemplified by Hill's work \cite{HillMoeckel}.  
That BB82 requires $\L = 0$ has been considered to be a prediction of the BB82 formulation as a 
separate theoretical entity from NM \cite{Pooley}; moreover $\L = 0$ appears to be true for the 
universe we are in.\fn{However, detailed cosmological study requires GR considerations.}  
Also, the possibility of formulating $\L \neq 0$ NM relationally should not be dismissed.  
While this goes against the Poincar\'{e} principle that Barbour advocates, 
that at most positions and velocities should need to be specified in RPM, 
this principle might just be a simplicity; for sure, $\L \neq 0$ is known to be much more 
complicated than $\L = 0$ by the work of Dziobek \cite{Dziobek} and of Poincar\'{e} \cite{Poincare}.  

Finally, there are some robustness issues.  
Can branches of physics external to particle mechanics be incorporated?  
Doing so would remove interference by external torques violating AM conservation, and the possibility 
of absolute space or time being bestowed upon particle mechanics by other branches of physics.  
Electromagnetism can probably be incorporated (see also p. 14). 
%
Dissipative processes such as linear air resistance and diffusion I can accommodate by a 
counterbalancing trick \cite{MF} that adjoins nonphysical fields (which may be regarded 
as `elsewhere') into the Lagrangian, but I cannot see any means of extending this to nonlinear 
dissipative processes such as air resistance quadratic in the velocity.   
In this paper, geared toward quantization, I stay clear of dissipative processes, 
which one would hope are in any case recastable in terms of more fundamental processes.  
What is excluded by considering only conservative dynamics? 
Is homogeneous quadraticity restrictive?  
While BS say ``pretty well all dynamical systems in both normal mechanics and SR field theory can 
be cast in such forms", both Lanczos \cite{Lanczos} and I \cite{Vanderson} have argued contrarily.  
In mechanics, the direct study of systems cast with linear `gyroscopic' terms would be precluded; 
linear terms also arise in the field-theoretic counterpart of this study for moving charges and for 
spin-1/2 fermions.  
While in principle nothing goes wrong if the treatment is extended to these, 
practical difficulties can arise if the ensuing complications thus brought into the actions 
are extended to their logical conclusion.  
However, as a compromise, as far as I know, nonhomogeneous quadraticity suffices to describe all 
established physics while retaining a form that is algebraically manageable,\fn{This 
nonhomogeneity does however spoil Barbour's assertion that all energies in the universe contribute 
to the emergent lapse quantity 
$\dot{O} \equiv \sqrt{\ssv_{\mbox{\tiny Total}}/(\sse - \sst_{\mbox{\tiny Total}})}$, 
which is related to his notion of perfect clock.}    
which gives Jacobi actions of the type  
\be
\sS = \int \d \lambda
\left( 
\sqrt{    (\se - \sv_{\mbox{\scriptsize Total}})
           \st_{\mbox{\scriptsize Quadratic}}     } 
+           \st_{\mbox{\scriptsize Linear}}          
\right)
\mbox{ } .
\ee

\section{Setting up the quantization of relational particle models}

I next address what kind of QM follows from such a formulation for the simplest models: 
$N = 3$ conservative, homogeneous quadratic $\L = 0$  models.  
In $d$ = 1 the configuration spaces are flat, so one can hope to employ standard mathematics.    
As discussed in Sec 4, I choose for the moment to continue to work with now only 
partly-reduced flat configuration spaces when passing to $d = 2$ and $3$. 
This is possible because of the good fortune that Jacobi coordinates 
preserve the separation of $\L = 0$.  

Making use of the position representation $\hat{\q}_{A} = {\q}_{A}$ and 
$\hat{\p}_{A} = -i\hbar\pa/\pa{\q}_{A}$, and denoting $\pa^2/\pa \q_I^2$ by $\triangle_{q_I}$     
the constraints of Sec 2 become the quantum constraints 
\be
\hat{\underline{\cal P}}\Psi(\q_A) \equiv \sum_{I=1}^N \frac{\pa}{\pa\q_I}\Psi(\q_A) = 0
\mbox{ } , 
\label{bbqc1} 
\ee
\be
\hat{\Q}\Psi(\q_A) \equiv 
\left(
\sum_{I=1}^N-\frac{\hbar^2}{2m_I}\triangle_{q_I} + \sv(\q_A) - \se 
\right) 
\Psi(\q_A) = 0
\mbox{ } .  
\label{bbqc2}
\ee
In 1-$d$, that is all, while for $d$ = 2 or 3 there is also (the portion relevant in dimension $d$ of)  
\be
\hat{\L}\Psi(\q_A) \equiv \sum_{I = 1}^N \q_I \mbox{\scriptsize $\times$} \frac{\pa}{\pa \q_I}\Psi(\q_A) = 0 
\mbox{ }.  
\ee
Note that there are no operator ordering ambiguities in any of these constraints (only in the case 
of the zero AM constraint are there products of $p$'s and $q$'s, and even there the order is 
unambiguous by symmetry--antisymmetry).  
I am also `lucky' in Dirac's sense \cite{Dirac}: 
the full $d$-dimensional set of constraints quantum-closes; 
moreover this is in direct parallel with the classical closure 
(i.e $\{\mbox{ } , \mbox{ } \} \longrightarrow \frac{1}{i\hbar}[ \mbox{ } , \mbox{ } ]$), 
so I do not present it.


The kinetic term in (\ref{bbqc2}) contains a sum of Laplacians on 
$\mbox{{\Large $\times$}}_{I = 1}^{N}\Re^{\md}$ in absolute, 
particle-position Cartesian coordinates $\q_I$. 
Toward addressing relational problems, I recast this in terms of relative coordinates.  
I wish the Laplacian to remain diagonal so as to exploit separability, 
so I employ relative Jacobi coordinates $\R_i$.   
The momentum constraint then becomes 
\be
\frac{\pa\Psi}{\pa\N} = 0 \mbox{ } , 
\label{37}
\ee
which expresses the translation-invariance of the wavefunction even more manifestly than (\ref{bbqc1}).  
Alternatively, one could start in relative Jacobi coordinates, in which case (\ref{37}) would 
have been built in automatically.  By either route, the remaining constraints are  
\be
\se\Psi = \sum_{i = 1}^{N - 1}\frac{-\hbar^2}{2\bar{m}}\triangle_{R_i}\Psi + \sv\Psi 
\ee
and (the portion relevant in dimension $d$ of)
\be
\hat{\L}\Psi \equiv 
\sum_{i = 1}^{N - 1} \R_i \mbox{\scriptsize $\times$} \frac{\pa}{\pa \R_i} \Psi = 0 \mbox{ } .
\ee
Thus I end up on the quotient space $Q/$\{Translations\}= $\mbox{\Large $\times$}_{i = 1}^{N - 1}\Re^{\md}$ in 
Cartesian coordinates $\R_i$.


For $N = 3$, I employ the Jacobi coordinates 
\be
\R_1 = \r_{12} =  \q_1 - \q_2 \mbox{ } ,
\label{aljac1}
\ee
\be
\R_2 =  {\cal N}
\left(
(m_1 + m_2)\q_3 - {m_1\q_1 - m_2\q_2}
\right)  
\mbox{ } ,
\label{aljac2}
\ee
\be
\R_3       = \q_3 \mbox{ } ;
\label{aljac3}
\ee  
(see Fig 1 for their meaning) with an as-yet unfixed normalization ${\cal N}$ for convenience.    
\begin{figure}[h] 
\centerline{\def\epsfsize#1#2{0.4#1}\epsffile{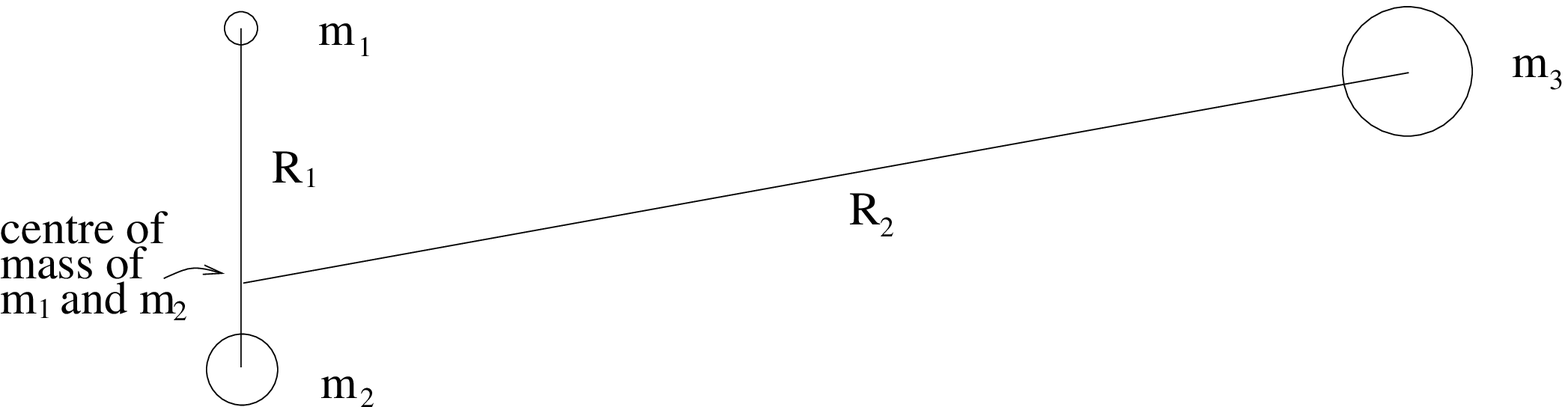}}
\caption[]{\label{TO3.ps}}\noindent{\footnotesize The meaning of (the standard normalization of) 
the 3-body problem relative Jacobi coordinates $R_1$ and $R_2$.  
Among other things, Jacobi coordinates are well-suited to study the effect of an extra mass $m_3$ 
on a well-understood subsystem $m_1$, $m_2$.}
\end{figure}

Then 
\be
\st \Psi = -\frac{\hbar^2}{2\mu_{12}}
\left(
\triangle_1 + \triangle_2
\right)
\Psi(\R_1, \R_2)
\ee
for $\mu_{12}$ the standard definition of reduced mass for a $m_1$, $m_2$ pair (I henceforth 
drop the $\mbox{}_{12}$ indices) and the choice of ${\cal N} = \sqrt{m_3/m_1m_2M}$.
Thus one has 2$d$-dimensional flat space problems, which look like standard 
time-independent Schr\"{o}dinger equations (TISE's), 
except that it is now the relative coordinates $\R_1$ and $\R_2$ and not the particle 
positions which play the role of Cartesian coordinates on the reduced configuration space, 
(and, for $d > 1$, there is an additional equation restricting the total AM to be zero).


I am principally interested in physically-motivated examples such as 
free, (piecewise) constant, harmonic oscillator (HO) and Newton--Coulomb type potentials.     
Furthermore, as each potential involved is expressible as a function of the relative separations 
$|\r_{IJ}|$ alone, choosing to use these potentials is a relational input, as befits the 
current investigation.  
This inherent relationalism in many commonly studied potentials is an early indicator of some 
mathematical similarities\fn{These similarities arise from the close (but conceptually distinct) 
parallel between the absolutist `intuitively' working in the barycentre frame since the barycentre 
motion is `mathematically uninteresting' and the relationalist holding the barycentre motion for a 
whole system/universe to be a fortiori {\sl meaningless}.}  
with the standard absolutist QM.
This is worth pointing out, since it means that, at least to start off 
with, one does {\sl not} get the wrong sort of formal mathematics by quantizing absolute rather 
than relational particle mechanics. 
Interpretational and mathematical differences appear later in the study.   
  
For such potentials, I demonstrate that passing to the relative Jacobi coordinates $\R_1, \R_2$ often 
permits separation 
\be
\Psi(\R_1, \R_2) = \psi_1(\R_1) \otimes \psi_2(\R_2)
\label{sepans}
\ee
{\sl into recognizable problems}.  
Indeed the $\q_I \longrightarrow \R_i$ map I employ is 
{\sl form-preserving} for many of these problems. 
I.e. it maps the 3$d$-dimensional configuration space quantum problem, for which the 
$\q_{I}$ ($I$ = 1 to 3) play the role of Cartesian coordinates and that is separable 
into 3 standard $d$-dimensional problems for each $\q_{I}$, 
to a corresponding $2d$-dimensional configuration space problem, 
for which $\R_i$ ($i$ = 1 to 2) play the role of Cartesian configuration 
space coordinates and that is separable into 2 of the {\sl same kind} of standard $d$-dimensional 
problems for each $\R_i$. 
The normalization of the 1-d models' wavefunctions is then standard and not required 
for this article's applications.

Thus my strategy is to solve relational quantum problems by setting up 
$\q_I \longrightarrow \R_i$ maps to standard separable problems, 
and then exploiting the well-established mathematical machinery for these.  
Only then do I consider, from a careful relationalist perspective, 
what the interpretation of the relational QM models is, and explore the new mathematics 
which comes upon considering these as whole systems/universes.

\section{Examples of $N$ = 3 relational QM}  

\subsection{Example 1: 1-d free problem} 

The free problem straightforwardly separates to 2 copies of\fn{In Sec 7 and 8, $i$ denotes a 
particular component rather than an index to be summed over.}  
\be
\frac{\hbar^2}{2\mu}\frac{d^2\psi_i}{dR_i^2} + \se_i\psi_i = 0 
\mbox{ } ,
\ee
giving wavefunctions 
$
\psi_i = e^{\pm i \sqrt{2\mu\sse_i}R_i/\hbar}
$
for a positive continuous spectrum.
The solution of the relational problem may then be reassembled as 
\be
\Psi_{\mbox{\scriptsize{\sffamily E}}} = e^{\pm i\sqrt\mu
\left( 
\sqrt{\sse_1} - \sqrt{2 \sse_2}{\cal N}
\right)
r_{12}/\hbar  } 
e^{\mp i\sqrt{2\mu \sse_2}
{\cal N}(m_1 + m_2)r_{23}/\hbar  } \mbox{ } . 
\ee
Finally, the unusual relational feature is that $\se_1$ and $\se_2$ are not independent: 
$\se_1 + \se_2 = \un$.

\subsection{Example 2: 1-d harmonic oscillators} 

For the single HO relational problem I exploit the following more widely applicable  
\noindent Move 1: as $\r_{12} = \R_1$, any potential depending at most on $\r_{12}$ is 
separable in $\R_1, \R_2$ coordinates, with identity map between the potential coefficients 
in $\q_I$-space and $\R_i$-space.  Thus 
\be
-\frac{\hbar^2}{2\mu}(\triangle_1 + \triangle_2)\Psi + \sv(|\R_1| \mbox{ alone })\Psi = \se\Psi 
\ee
\be
\Rightarrow \mbox{ } -\frac{\hbar^2}{2\mu}\triangle_1\psi_1 + \sv(|\R_1|)\psi_1 = \se_1\psi_1 
\mbox{ } , \mbox{ }
-\frac{\hbar^2}{2\mu}\triangle_2\psi_2 +  = \se_2\psi_2  \mbox{ } \mbox{ for }  
\se_1  + \se_2 = \un 
\mbox{ } .  
\label{Etie}
\ee
Thus, let the HO be without loss of generality between 
particles 1 and 2 so that $R_1$ follows the separated-out 1-d problem 
\be
\frac{\hbar^2}{2\mu}\frac{d^2\psi_i}{dR_i} - \frac{H_iR_i^2}{2}\psi_i + {\se}_i\psi_i = 0 \mbox{ } .  
\ee
This is solved by wavefunctions 
$
\psi_i(\nn_i) = \mbox{H}_{\mn_i}(Y_i)e^{-Y_i^2/2}  
$
for $Y_i = (\mu H_i/\hbar^2)^{1/4}R_i$ and H$_{\mn_i}$ the $\nn_i$th Hermite polynomial and 
corresponding spectrum $\se_i = \sqrt{H_i/\mu}\hbar\left(\frac{1}{2} + \nn_i\right)$, 
$\nn_i\in N_0$.  While, $R_2$ follows the separated-out free problem of Subsec 7.1.  
Thus
\be
\Psi_{\mn} = 
\left( 
e^{\pm\sqrt{2\mu \sse_2}{\cal N}(m_1 + m_2)r_{23}/\hbar} 
\right)
\left(
e^{      \pm i\sqrt{2\mu \sse_2}{\cal N}m_1 r_{12}/\hbar - \sqrt{\mu H_1}r_{12}^2/2\hbar    }
\mbox{H}_{\mn}
\left(
(\mu H_1/\hbar^2)^{1/4}r_{12}
\right)
\right) 
\mbox{ } .
\ee
Finally, $\nn \equiv \nn_1$ and $\se_2$ aren't independent: 
$\sqrt{    {H_1}/{\mu}    }\hbar    \left(    \nn +\frac{1}{2}    \right)  + \se_2 = \un$.  

For 2 or 3 HO potentials, map according to the following Move 2.  
Denoting the numerical coefficient in the $IJ$ pair's potential by $h_{IJ}$ 
and the ith relative coordinate's potential coefficient by $H_i$, 
for 2 HO's, without loss of generality the potential is  
\be
h_{23}\r_{12}^2 + h_{13}\r_{13}^2 = \frac{1}{(m_1 + m_2)^2}
\left(
\left(
h_{23}m_1^2 + h_{13}m_2^2
\right)
\R_1^2
+ \frac{h_{23} + h_{13}}{{\cal N}^2}\R_2^2 
+ \frac{m_1h_{23} - m_2h_{13}}{{\cal N}}2\R_1\cdot\R_2
\right)
\mbox{ } ,  
\ee
so there is the separability restriction\fn{These large families serve to provide simple 
examples in this paper.  Completely general 2 and 3 HO set-ups can be accommodated by rotating the 
Jacobi coordinates.} 
${h_{13}}/{h_{23}} = {m_1}/{m_2}$,  
and the relations between the $\q_I$ and $\R_i$ HO's Hooke coefficients are
$H_1 = {h_{23}\mu}/{m_2}$, $H_2 = {h_{23}\mu M}/{m_2m_3}$.
For 3 HO's, the same separability restriction holds again and the relations are
$H_1 = h_{12} + {h_{23}\mu}/{m_2}$, $H_2 = h_{23}\mu M/m_2m_3$.
Thus one obtains two separated-out 1-$d$ HO problems, 
with coefficients in each case as given above.    
The desired solution in relative separation coordinates is then   
\be
\Psi_{\mn} = 
\mbox{H}_{\mn_1}
\left(
(\mu H_1/\hbar^2)^{1/4}r_{12}
\right)
\mbox{H}_{\mn_2}
\left(
(\mu H_1/\hbar^2)^{1/4}{\cal N}(m_1r_{13} + m_2r_{23})
\right)
e^{-
\sqrt{\mu H_1}r_{12}^2/\hbar +  \sqrt{\mu H_2}\left({\cal N}^2(m_1 + m_2)r_{23} + m_1r_{12}\right)^2/\hbar} 
\mbox{ } .  
\ee 
Finally, $\nn_1$ and $\nn_2$ aren't independent: 
$\sqrt{    {H_1}/{\mu}    }\hbar\left(\nn_1 +\frac{1}{2}\right)  + 
 \sqrt{    {H_2}/{\mu}    }\hbar\left(\nn_2 +\frac{1}{2}\right)  = \un$.  
Notice how the 2 and 3 HO solutions are twisted rather than trivial 
when expressed in terms of $r_{IJ}$ coordinates.

\subsection{Example 3: a simple atomic model in 3-d} 

This is motivated 1) for more accurate real-world modelling purposes.
2) 
To tackle rotation/AM, which is the interesting and 
complicated part of relationalism, and the inclusion of which gives        
toy models that resemble geometrodynamics more closely (see e.g. {\bf II}), 
such as the $N = 3$, $d = 3$ `Triangle Land' that Barbour's speculations \cite{EOT} are based on

As noted on p.7 
AM doesn't spoil separability.  
It is furthermore significant that the AM in $q_I$ space is mapped to its mathematical analogue in 
$R_i$ space, so that one can treat AM there as one usually does in $\q_I$ space. 
Full polar coordinates (i.e polars in 2-$d$ and spherical polars in 3-$d$) for each $\R_i$ 
are useful in this respect.  

For a single Coulomb potential in 3-d (representing a hydrogen atom together with a free 
neutral particle), one immediately has separability by Move 1, into the 3-d Helmholtz equation 
and the simple atomic model.
Additionally, the AM are tied between the 2 problems, so spherical polars 
$(\rho_i, \theta_i, \phi_i)$ are well-suited.   
Then both of these building blocks have the angular part $\YY_{\ml_i\mm_i}(\theta_i, \phi_i)$ 
(spherical harmonics).

The radial part of the separated-out free problem is solved by the spherical Bessel functions 
$
\psi_{\rho_i}  = \jj_{\ml_i}(k_i \rho_i) \propto \JJ_{\ml_i + {1/2}}(k_i \rho_i)/{\sqrt{\rho_i}},   
$ 
where the $k_i$ are continuous positive values related to $\se_i$ by $k= \sqrt{2\mu\se_i}/\hbar$.   
The radial part of the separated-out simple atomic problem is solved in terms of 
associated Laguerre polynomials 

\noindent
$
\psi_{\rho_i}  = \rho_1^{\ml}\mbox{L}^{2\mn + 1}_{\mn - \ml - 1}
\left(
{\mu k}\rho_1/\hbar^2
\right)
e^{- \mu k \rho_1/\hbar}.  
$
The corresponding energy eigenspectrum of bound states is 
$\se_i = - \mu k/2\hbar^2\nn^2$ for an attractive potential 
$\sv = - k/\rho_1$.    
There is also a continuum of positive energies representing ionized states.    
Thus overall this simple relational atom model is solved by 
\be
\Psi_{\mn\ml\mm} = \YY_{\ml\mm}(\theta_1, \phi_1)
                      \rho_1^{\ml}\mbox{L}^{2\mn + 1}_{\mn - \ml - 1}({\mu k}\rho_1/\hbar^2)
                      e^{-\mu k \rho_1/\hbar^2} 
                      \YY_{\ml -\mm}(\theta_2, \phi_2)j_{\ml}
(\sqrt{2\mu\se_i}\rho_2/\hbar)
\ee
where $\se_1$ and $\se_2$ aren't independent: $\se_1 + \se_2 = \un$ 
and the angular momentum conterbalancing explained on p.13 is in use.  
I don't present this here re-expressed in terms of $\R_i$ and then $r_{IJ}$ as that becomes 
messy.  

As regards further examples, Move 1 is widely applicable, while Move 2 also 
works for 2 or 3 isotropic HO's in 2- or 3-d (see \cite{v2} for these).

\section{Interpretation of N = 3 relational QM}

\subsection{Decent semiclassicality}       

While one common requirement is for the wavelength $\lambda_{\mbox{\scriptsize Q}}$ to be smaller 
than some characteristic scale $l_{\mbox{\scriptsize C}}$ of the problem, 
there is no universal rigorous notion of semiclassical limit for quantum theories.  
Here are various procedures that one might consider in connection with such an investigation, 
all of which play some role in papers {\bf I} or {\bf II}.  

\noindent 1) Consider the spread (i.e width) of the wavefunctions 
[this is essentially what BS do].    

\noindent 2) Furthermore investigate how localized wavepackets are as a whole 
(as opposed to the spread of each wavefunction in their summand/integrand).

\noindent 3) Consider what happens to the system for large quantum numbers whereupon 
the wavefunction becomes wiggly on scales much shorter than $l_{\mbox{\scriptsize C}}$.  

\noindent 4) Consider a WKB ansatz for the wavefunction and expand in powers of 
$\lambda_{\mbox{\scriptsize Q}}/l_{\mbox{\scriptsize C}}$.  

\mbox{ } 

\noindent{\bf Spread of the wavefunctions}

\mbox{ }

\noindent Barbour and Smolin's first objection BS1 was to the sensitivity of the spread of the 
wavefunctions of large-mass particles to the values of the mass of small-mass particles.  
This was in connection with piecewise-constant potential models 
with two small masses and one large one.  
I clarify why this is not a problem as follows.  
First, bear in mind that in the absolutist quantization, one thinks primarily of particle wavepackets, 
but in a 1-d relationalist quantization one should think of {\sl relative distance wavepackets}.      
Then, intuitively, once one rephrases one's standard quantum intuition about small 
masses being more spread out in relational terms, it is clear that the position uncertainty 
of the small mass dominates the relational formulation's relative separation 
uncertainty between that small mass and a large mass (Fig 2).  
\begin{figure}[h] 
\centerline{\def\epsfsize#1#2{0.4#1}\epsffile{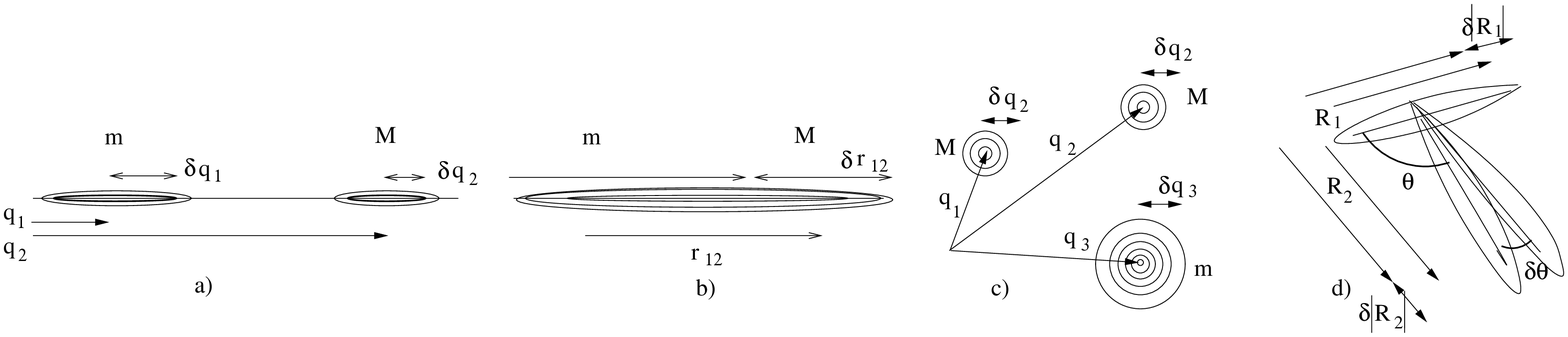}}
\caption[]{\label{TO4.ps}}\noindent{\footnotesize For two particles, position spread a) versus relative 
separation spread b).  For 3 particles in $d > 1$, c) and d) are position spread versus 
a choice to represent relative spread in Jacobi coordinates [$\theta$ = arccos($\R_1\cdot\R_2/|\R_1||\R_2|$)].}
\end{figure}

BS's specific example is akin to my Example 1, except that they  
have a constant $U$ and set $\un = 0$ while I have no $U$ but a constant $\un$.
Converting, the wavefunction may then be rewritten as 
\be
\Psi = e^{        i\sqrt{|U|}      
\left(      
r_{12}
\left(    
\frac{sin\theta}{\sqrt{\nu_{13}}} - \frac{n_3cos\theta}{\sqrt{\chi}}    
\right) 
+ r_{23}
\left(    
\frac{sin\theta}{\sqrt{\nu_{13}}} + \frac{n_1cos\theta}{\sqrt{\chi}}      
\right)          
\right)/\hbar }
\ee
for $n_a = 1/m_a$, $\nu_{ab} = 1/\mu_{ab}$, 
$\chi = n_3^2\mu_{12} + n_1^2\mu_{23} + 2n_1n_2n_3$ and where the mass-independent 
constants present are parametrized by an angle $\theta$.  
This has much nicer 1,3 symmetry than 1,2 symmetry, so in setting two 
of the masses to be equal to match BS's example, I opt for $m_1, m_3 = m << M = m_2$  
Then the wavefunction goes as 
\be
\Psi \mbox{ } \sim \mbox{ } e^{iJ\sqrt{m}r_{12}/\hbar}e^{iK\sqrt{m}r_{23}/\hbar}
\ee
(for $J$, $K$ mass-independent constants).   
Thus, indeed as BS claim, the small masses dominate all the uncertainties.    
But by my above interpretation, these uncertainties are actually in the 
separations between a big mass and a small mass, so this situation conforms 
to standard quantum intuitions rather than constituting an impasse.  
The {\sl truly} relevant test to establish whether there is a semiclassical limit problem 
is rather to check what happens if  $m_1, m_3 = M >> m = m_2$, for then there is a big 
mass--big mass relative separation, and it is {\sl this} which one would not 
expect to be influenced much by a small mass somewhere else.  
And indeed, upon performing the new approximation, and isolating the big mass--big 
mass separation $r_{13}$ as the variable whose spread is of relevance, I find that 
\be
\Psi \mbox{ } \sim \mbox{ } e^{ i \bar{J}\sqrt{{M}} r_{13}/\hbar}e^{-i\bar{K}\frac{m}{\sqrt{M}}r_{12}/\hbar}
\ee
(for $\bar{J}$, $\bar{K}$ mass-independent constants) 
so that indeed only the big masses contribute significantly to the spread in the 
big mass-big mass separation.  
This basic conclusion is unaffected by having the two identical masses replaced by merely similar 
masses, and holds widely throughout the models presented in this paper when suitable pairs of 
quantities are set to be relatively large and small.  

\mbox{ }

\noindent{\bf Wavepackets} 

\mbox{ } 

\noindent Consider the 1-$d$ problems which separate out of the relational problems of this section, 
piecemeal and as formal pieces of mathematics in which an eigenspectrum and wavefunctions are 
obtained from a differential equation and boundary conditions that depend on some abstract set 
of variables and parameters.  
Then the wavepackets built up by summing and/or integrating the wavefunctions (generally with 
some weighting) over the eigenspectrum may also be considered as formal pieces of mathematics.  
My first point is that, at the level of the separated-out pieces of the relational problems 
considered in this section, this formal mathematics is the same as in the usual absolutist 
quantization. 
Thus the piecemeal construction of wavepackets for the 1-d quantum problems does not care whether 
these arise from separation in relational problems or in absolutist ones, so both behave equally well.    
The best-known examples of these wavepackets are the fixed-size one for the free particle and 
the pulsating one for the harmonic oscillator (see e.g. \cite{Schiff, Robinett}).      

A first difference between wavepackets in the absolutist and relational QM schemes 
arises in their interpretation.  
This parallels the above situation with the spreads.
A second difference is that there are limitations building composite wavepackets in the relational case.  
Unlike in the absolutist case, composition of subsystem wavepackets cannot be 
extended to include the whole system.  
This is due to the energy of the universe being a fixed quantity.    

Also note that the application of polar coordinates for $d > 1$ brings out that the 
standard QM interpretation is close to being relational.  
This is most familiar in the study of the hydrogen atom, for which   
a simple standard approach is to treat the proton as fixed and then consider the spread of the 
radial separation $\rho$ between the proton (or more accurately the atom's barycentre) and the electron. 
All that is missing as regards obtaining a fully relational perspective is to consider 
the position of the barycentre not only to be uninteresting but also to be meaningless.  
Then one considers the spread in $|\R_1| (= |\r_{12}| \equiv \rho_1)$.  
The ready availability of this familiar picture is one reason why it is unfortunate 
that BS restricted their study to 1-$d$ examples.  
I should add that interpreting $d > 1$ RPM's furthermore 
requires an additional notion of {\sl spread in relative angle} [Fig 2 d)].

\subsection{\bf Interesting features of the RPM examples as closed universe systems} 

\noindent{\bf Subsystem energy interlocking and truncation, gaps in the energy spectrum}

\mbox{ }

\noindent 1) As mentioned above, there is energy interlocking between constituent subsystems.  
E.g. this requires the above 
na\"{\i}vely free problem to have a line segment rather than a quadrant as its overall eigenspectrum, 
and the single HO and free particle system to have a set of points rather than an infinity of lines 
as its overall eigenspectrum, and the coupled HO's have a small set of 
points rather than a regular lattice as its eigenspectrum (Fig 3).  
\begin{figure}[h] 
\centerline{\def\epsfsize#1#2{0.4#1}\epsffile{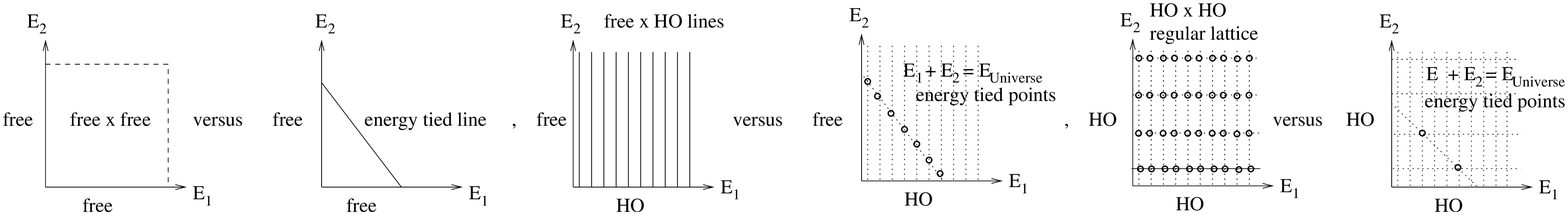}}
\caption[]{\label{TO5.ps}}\noindent{\footnotesize Effect on spectra 
of $\un$ taking a forever fixed value.}
\end{figure}
By energy interlocking the small universe models built up from individual problems' wavepackets 
additionally contain a delta function 
\be
\delta
\left(\sum_{\Delta \in \mbox{ subsystems}} \se_{\Delta} - \un
\right)
\ee
acting inside the sums and integrals required to build it up, which causes it to differ 
mathematically from e.g the direct product of subsystem wavepackets in a fully separable universe.  

Moreover, unlike in the usual interpretation of few-particle QM, 
the energy here is the energy of the universe, which is not only fixed, but is also  
a separate attribute of the universe so that the fixed value it takes need bear no relation 
to the eigenspectra of the universe's contents.  
This leads to the following effects.

\noindent 2) In my multiple HO example, all the energies [$\se_1(\nn_1)$ and $\se_2(\nn_2)$] 
are positive.  Then $\un < \se_1(0) + \se_2$ so no wavefunctions exist.  
Universes failing to meet the zero point energy of their content fail to have a wavefunction.  

\noindent 3) Such universes could rather fail to meet the energy required for just some of 
the states which lie above a given energy that is greater than the zero point energy.  
then one would obtain not the conventional eigenspectrum but rather a {\sl truncation} of it.  

\noindent 4) In my 2 or 3 HO examples, one is required to solve $k_1\nn_1 + k_2\nn_2 = q$ for 
$\nn_i \in |N_0$, $k_i = \sqrt{  \frac{H_i}{\mu}}\hbar $ and 
$q = \un - \frac{\bar{h}}{2}\frac{\sqrt{H_1} + \sqrt{H_2}}{\sqrt{\mu}}$.  
Then if $k_i$ and $q$ mismatch through some being rational and some irrational, or even if all are 
integers but the highest common factor of $k_1$ and $k_2$ does not divide $q$, no solutions exist.  
This can also be set up to give {\sl gaps} in what would otherwise look like a 
truncation of the conventional eigenspectrum.  
Similar effects can be achieved by more complicated matches and mismatches in the other examples' 
$\se_i$ dependences on $\nn_i$.    

\mbox{ }

\noindent{\bf How energy interlocking does not affect the `recovery of reality' in practice for large universes }

\mbox{ } 

\noindent Objection BS2 is due to a less developed version of the above material.  
However, I develop the following conditional counterstatements.  

\noindent 5) In universes which furthermore contain free particles, 
because these have continuous spectra, 
the missing out of some conventional states by 4) cannot occur.  
This is the case e.g for the 1 HO example above, 
while `tensoring' free particles with the 2 or 3 HO setting above 
alleviates this `missing state' problem 
in the setting of a universe with a slightly larger particle number.    
 
\noindent 6) If negative energies are possible, 
then subsystems can attain energies higher than $\un$. 
So if one tensors a negative potential example such as hydrogen with an independent HO pair, 
one can have less truncation of the HO.  
One will still have many, or all, states missing, 
depending on how the coefficients of the two independent subsystems' potentials are related.  
But if one then tensors in free particles one can have everything up to the truncation.  
Thus truncation can be displaced at least for some universes, by a modest increase in 
constituent particle number.  
It should be noted that there is a mismatch between e.g. the HO which has 
excited states unbounded from above in conventional QM and hydrogen  
which has negative energy states bounded from below.  
I get out of this difficulty by pointing out that in any case very high positive energy states are 
unlikely to be physically meaningful; at the very least the physical validity of the model 
would break down due to e.g pair production and ultimately the breakdown of spacetime.  
To ensure one's model can attain high enough (but finite) positive energies, wells that are deep 
and/or numerous enough can be brought in.    

7) Unlike 2)--4), energy interlocking {\sl does not} go away with particle 
number increase or the accommodation of a variety of potentials within one's universe model.
However, the large particle number and high quantum number aspects of 
semiclassicality are relevant here.  
Subsystems remain well-behaved and all experimental studies in practice involve subsystems.  
But subsystems may be taken to have conventional wavepackets insofar as these {\sl are} products of 
their constituent separated-out problems' wavepackets.  
It is still true that these may be truncated rather than built out of arbitrarily many eigenfunctions 
along the lines of 1) to 3) above, but this will be alleviable in practice by ensuring that 
sufficient additional free particles or particles whose mutual potentials are of an opposing sign to 
the original subsystem's.  
In such a framework, the correlation of a quantity within a subsystem with another in 
the rest of the universe is overwhelmingly likely to evade detection provided that 
the rest of the universe contains plenty of other particles.  
And of course the real universe is indeed well-populated with particles.  

Thus, BS's second bad semiclassicality objection should be replaced by:  
1--4) lead to simple small relational particle universes constituting 
good toy models for gaining a deeper closed universes and of the Problem of Time, 
while 5--7) ensure that thinking of the (diverse, large particle number) real universe in terms of  
relational particles is {\sl not} in practice compromised by any manifest bad semiclassicality.
I should also caution that solving TISE's is not as complete as setting up the full machinery of QM 
so more detailed contentions may emerge in further work.

Whether the WKB ansatz may additionally be applied in to whole universes is an issue relevant 
to the semiclassical approach to quantum cosmology in the general (rather than just RPM) 
context.  I touch on this issue in Sec 8.3, but mostly leave its discussion to {\bf II}.8 and \cite{Soland}. 
For the moment I show that increasing the dimension away from 1 reveals further such `small closed 
universe' effects.  

\mbox{ }

\noindent{\bf Angular momentum counterbalancing}

\mbox{ }

\noindent
2-$d$ RPM solutions depend on 2 and not 4 quantum numbers via the AM constraint as follows: 
$L_i\psi_i = \nm_i\psi_i$ $i = 1, 2$ arise in each separated problem, but these are 
`joined together' by 
$0 = \hat{L}\Psi = \hat{L}_1 \otimes \hat{1}_2 \Psi + \hat{1}_1 \otimes \hat{L}_2\Psi 
   = \nm_1\Psi + \nm_2\Psi $, so $\nm_1 = - \nm_2 \equiv m$.  
3-$d$ RPM solutions such as Example 3 depend on 3 and not 6 quantum numbers via a {\sl more elaborate} 
angular momentum counterbalancing as follows: 
$\nm_1 = - \nm_2 \equiv \nm$ as for the 2-d case using $0 = \hat{L}_z\Psi$ in place of 
$0 = \hat{L}\Psi$.  
But now, also, $0 = \hat{L}_x\Psi$ and $0 = \hat{L}_y\Psi$, so 
$\hat{L}_{x1}\Psi = - \hat{L}_{x2}\Psi$ 
and $\hat{L}_{x1}\Psi = - \hat{L}_{x2}\Psi$.
Thus, using $\hat{L}_1^2\psi_i = \nl_i(\nl_i+1)\psi_i 
\mbox{ } \Rightarrow \mbox{ } 
(\hat{L}_{xi}^2 + \hat{L}_{yi}^2)\psi_i = (\nl_i(\nl_i + 1) - \nm_i^2)\psi_i$ for each of 
$i = 1, \mbox{ } 2$, so $\nl_1(\nl_1 + 1) - \nm_1^2 = \nl_2(\nl_2 + 1) -\nm_2^2$.
But $\nm_1 = -\nm_2 \equiv \nm$ and $\nl_1, \nl_2 \in {\cal N}_0$ so there is no choice but 
$\nl_1 = \nl_2 \equiv \nl$.
Thus whole universe models have a variety of additional na\"{\i}vely unexpected 
correlations between their ordinary-looking constituent subsystems due to limited 
energy resources and due to having to balance out each others' angular momentum.
A similar argument to that presented for energy interlocking leads to one expecting 
angular momentum counterbalancing not to be noticed in the study of subsystems within a large universe.  

\mbox{ }

\noindent{\bf Further types of question about the `recovery of reality'}

\mbox{ }

\noindent Conceptualizing in terms of an atom and some other subsystem, 
lends itself to raising further types of questions.  
It is all very well that these subsystems can be constructed to 
have the right sort of spectra in the sense of energy levels (mathematical sets of eigenvalues).  
But are {\sl transitions} between these, most obviously manifested by spectra (in the distinct, 
practical sense of emissive/absorptive frequency patterns) or chemical reactions, possible?  
After all, the universes in question are governed by TISE's which possess solely 
stationary solutions?  
One answer, developed in {\bf II}.7 and \cite{Soland}, is that this set-up {\sl is} nevertheless 
capable of allowing {\sl subsystems} to take on {\sl the appearance} of dynamics.  
This is a {\it semiclassical approach} which relies on the WKB ansatz 
and on (perhaps small) terms that cause nonseparability.  
One situation which has nonseparability is the multi-Coulomb 
potential (in some cases through multiple charged particles, while all models 
will have small gravitational interactions). 
Consistent records schemes such as Barbour's concern a different answer  
(see {\bf II}.8.  for further comparison of these).  
N.B. that, while this discussion of apparent dynamics from TISE's may appear unusual at first sight, 
it {\sl is} a simplified discussion about the plausible situation that our own universe, 
which certainly possesses subsystems which appear to be dynamical, 
may (at some level) be describable {\sl overall} by a GR TISE: 
the so-called Wheeler--DeWitt equation (See {\bf II} for references). 

It is subsequently relevant to ask about {\sl mechanism}.  
Insisting on interpreting such models as closed (i.e self-contained) amounts to {\sl not} assuming 
a `surrounding photon sea'.  
Does the absense of this preclude all dynamical processes?  
The answer is no, insofar that further particles of the same species can also mediate energy at some 
effective level (this holds as far as Fermi theory, overcoming the non-renormalizability of which 
leads to quantum electrodynamics, in which photons {\sl are} then necessary for interactions). 
For atoms, it is well known that a grazing free particle can result in electron 
excitation, in ionization or in the (4-particle) Auger effect \cite{Mott}.  
If one must furthermore deploy a counterbalancing particle/subsystem in order to recover standard 
results, at least 4 or 5 body problems are required to model such situations.   

A separate point is that, while photon-free transitions as above are known, many {\sl more} effects 
require interaction by photons or taking into account the atomic \underline{E} and \underline{B} fields.  
This is one reason why it would be interesting to extend the present work so as to `build in' 
electromagnetism by considering closed relational Newton--Maxwell universes.  
This would also serve both as a robustness check for the results of paper {\bf II} and as a toy 
model for closed Einstein--Maxwell universes.      
Inclusion of intrinsic spin would also be useful.

\section{Conclusion}

As regards the classical absolute versus relative motion debate, 
I have provided a concrete synthesis of directly-formulated relational particle mechanics (RPM)    
which is equivalent to a reformulation of the portion of Newtonian mechanics that is 
conservative, has $\L = 0$ and whose kinetic term is homogeneous quadratic in its velocities.  
I have presented this in relative Lagrange variables and in relative Jacobi variables.  
I have furthermore presented improved (counterbalancing subsystem) arguments 
that adopting this portion or the straightforward nonhomogeneous enlargement of it 
is not for many purposes a major restriction at the classical level.  

As regards QM implications of adopting the relational stance, 
I have been able to go further than in previous relational studies by bringing in Jacobi variables.  
Thus I have been able to work with more complicated, more realistic and more relationally motivated 
quantum RPM models than Barbour and Smolin (BS) or any other paper in the RPM  
literature to date.  
This is based on how many simple problems are separable in Jacobi coordinates 
and share much formal mathematical structure with the conventional absolutist approach.  
Thus at the very simplest level, the answer to my question about whether absolutism has misled QM 
is {\sl no}, in particular if one continues to adhere with conventional QM's emphasis 
(see e.g. \cite{DiracQM}) on formal structures and their manipulation.   
However, differences in the {\sl interpretation} of that mathematics are readily manifest  
due to the one employing particle positions and the other employing interparticle (cluster) 
separations.  
Indeed the bona fide relational interpretation that I provide enables me to reject 
BS's spread sensitivity objection to RPM models having a good semiclassical limit.  

Moreover the answer is {\sl yes as regards whole closed universes}.  
The models of this paper exhibit several interesting features,  
which suffice to show that insisting on modelling a closed universe as a whole courts difficulties 
well beyond those encountered in studying its constituent subsystems.  
(Of course, there is more to modelling quantum cosmology than just this.)
These features are energy interlocking, AM counterbalancing, eigenspectrum gaps due to the universe 
and its contents being mismatched, and eigenspectrum truncation due to finite resources.  
While these mathematical observations are an extension of BS's second objection to RPM models 
having a good semiclassical limit, moreover these effects do not affect the `recovery of reality' for 
a large universe with sufficiently varied contents.  
For, the presence of free particles and of potentials of both signs serve to overcome gaps and 
noticeable truncation, while the practicalities of experimentation involving only small subsystems 
means that interlocking and counterbalancing would be likely to go unnoticed.  

\mbox{ }

\noindent{\bf Acknowledgments}

\mbox{ }

\noindent 
I thank my Father, who taught and encouraged me prior to his death in 1995.  
I thank also my other early teachers, lecturers, supervisors and professors and the 
Royal Society of Chemistry for illusioning me.       
I thank Professor Malcolm MacCallum, Dr Julian Barbour and Professor Don Page 
for more recent guidance toward this project.  
I thank Professor Claus Kiefer for sending me the BS preprint while Julian was ill. 
I thank Professor Bruno Bertotti, Dr Martin Bojowald, Dr Harvey Brown, Mr. Brendan Foster, 
Dr Laszlo Gergely, Professor Gary Gibbons, Ms. Isabelle Herbauts, Dr Bryan Kelleher, 
Dr Adrian Kent, Professor Jacek Klinowski, Professor Niall \'{O} Murchadha, 
Dr Jonathan Oppenheim, Dr Oliver Pooley, Professor Reza Tavakol, Professor Lee Smolin, 
Dr Vardarajan Suneeta  and Dr Eric Woolgar for discussions.    
I thank the Barbour family, Peterhouse and DAMTP for hospitality 
during which some of this work was done.   
I acknowledge funding at various stages from Peterhouse and the Killam Foundation.  



\begin{thebibliography}{99}

\footnotesize

\bibitem{Newton}              I. Newton, {\it Philosophiae Naturalis Principia Mathematica} (1686).  
                              For an English translation, see e.g I.B. Cohen and A. Whitman (University of California Press, Berkeley, 1999). 
                              In particular, see the Scholium on absolute motion therein.  

\bibitem{L}                   See \it The Leibnitz--Clark Correspondence\normalfont, ed. H.G. Alexander (Manchester 1956).  
                              
\bibitem{B}                   Bishop G. Berkeley, \it The Principles of Human Knowledge \normalfont (1710);  
                 \it Concerning Motion (De Motu) \normalfont (1721). 
                              
\bibitem{M}                   E. Mach, {\it Die Mechanik in ihrer Entwickelung, Historisch-kritisch dargestellt} (J.A. Barth, Leipzig, 1883).  
                              The English translation is {\it The Science of Mechanics: A Critical and Historical Account of its Development} (Open Court, La Salle, Ill. 1960).    

\bibitem{Lag}                 J.-L. Lagrange, {\it Le probleme des trois corps, Oeuvres}, Vol 6, 229 (1772).  
                              
\bibitem{Jac}                 C.G.J. Jacobi, {\it Sur l'elimination des noeuds dans le probleme des trois corps}, Math Werke Vol. 1 30 (1843). 

\bibitem{Relper}              Relevant aspects of Lagrange and Jacobi's work are also discussed in e.g 
                              A. Chenciner, Ravello Summer School Notes (unpublished, from 1997);  
                                                            J. Barbour, material for unfinished volume II of {\sl Discovery of Dynamics}.      
                                                          
\bibitem{Dziobek}             O. Dziobek, {\it Die Mathematischen theorien der planeten-bewegungen }
                              (Barth, Leipzig 1888), available in English as 
                              {\it Mathematical Theories of Planetary 
                              Motions} (1892), now available as (Dover, New York 1962).

\bibitem{Poincare}            H. Poincar\'{e}, {\it Science et Hypotheses} (1902), 
                              English translation (Science Press, New York 1905).   


\bibitem{BB82}                J.B. Barbour and B. Bertotti, Proc. Roy. Soc. Lond. \bf A382 \normalfont 295 (1982).

\bibitem{BS}                  J.B. Barbour and L. Smolin, unpublished preprint dating from 1989.  

\bibitem{Rovelli}             C. Rovelli, p 292 in {\it Conceptual Problems of Quantum Gravity} ed. 
                              A. Ashtekar and J. Stachel (Birkh\"{a}user, Boston, 1991).

\bibitem{Smolin}              L. Smolin, in {\it Conceptual Problems of Quantum Gravity} ed. 
                              A. Ashtekar and J. Stachel (Birkh\"{a}user, Boston, 1991).

\bibitem{B94I}                J.B. Barbour, Class. Quantum Grav. \bf 11 \normalfont 2853 (1994).

\bibitem{B94II}               J.B. Barbour, Class. Quantum Grav. \bf 11 \normalfont 2875 (1994).

\bibitem{Buckets}             J.B. Barbour, {\it GR as a perfectly Machian theory}, in {\it Mach's principle: From Newton's Bucket to Quantum Gravity} ed. J.B. Barbour and H. Pfister (Birkh\"{a}user, Boston 1995).

\bibitem{LB}                  D. Lynden-Bell, in \it Mach's principle: From Newton's Bucket to Quantum Gravity\normalfont, ed. J.B. Barbour and H. Pfister (Birkh\"{a}user, Boston 1995).
   
\bibitem{EOT}                 J.B. Barbour, {\it The End of Time} (Oxford University Press, New York 1999).

\bibitem{Gergely}             L.\'{A} Gergely, Class. Quantum Grav. {\bf 17} 1949 (2000), gr-qc/0003064. 

\bibitem{GergelyMcKain}       L.\'{A} Gergely and M. McKain, Class. Quantum Grav. {\bf 17} 1963 (2000), gr-qc/0003065.

\bibitem{RWR}                 J.B. Barbour, B.Z. Foster and N. \'{O} Murchadha, Class. Quantum Grav. \bf 19 \normalfont 3217 (2002), gr-qc/0012089.

\bibitem{Kuchar92}            K.V. Kucha\v{r}, in \it Proceedings of the 4th Canadian Conference on 
                              General Relativity and Relativistic Astrophysics\normalfont,  
                              ed. G. Kunstatter, D. Vincent and J. Williams (World Scientific, Singapore, 1992).

\bibitem{Kiefer}              C. Kiefer, {\it Quantum Gravity} (Clarendon, Oxford 2004).

\bibitem{Landerson}           E. Anderson, in {\it General Relativity Research Trends, Horizons in World Physics} 
                              {\bf 249} Ed. A. Reimer (Nova, New York 2005), gr-qc/0405022.  

\bibitem{Barbourphil}         J.B. Barbour, in {\it Quantum concepts in space and time} ed. R. Penrose and C.J. Isham (Oxford University Press, Oxford, 1986).   
                              J. Earman, \it World Enough and Space-Time: Absolute versus Relational Theories of Space and Time \normalfont (MIT Press, Cambridge MA, 1989).
                              L. Smolin, in {\it Time and the Instant} ed.  R. Durie (Clinamen Press, Manchester 2000), gr-qc/0104097.
                              J.N. Butterfield, Brit. J. Phil. Sci. {\bf 53 } 289  (2002), gr-qc/0103055;    
                              J.B. Barbour, in {\it Decoherence and Entropy in Complex Systems (Proceedings of the Conference DICE, Piombino 2002} ed. H.-T. Elze (Springer Lecture Notes in Physics 2003),  gr-qc/0309089.  

\bibitem{Panderson}           These results were announced in outline in E. Anderson AIP Conf. Proc. {\bf 861} 285 (2006), gr-qc/0509054.    

\bibitem{Pooley}              O. Pooley and H.R. Brown, Brit. J. Phil. Sci. \bf 53  \normalfont 183 (2002); 
                              O. Pooley, Chronos (Proceedings of the Philosophy of Time Society 2003-4). 

\bibitem{PaperII}             E. Anderson  Class. Quant. Grav. {\bf 23} 2491 (2006), gr-qc/0511069.    

\bibitem{Einstein}            A. Einstein, Sitz. Preuss. Ak. Wiss. 142 (1917);                                 
                              {\it The Meaning of Relativity} (Princeton University Press, Princeton 1955).  

\bibitem{Battelle}            J.A. Wheeler, in \it {Battelle rencontres: 1967 lectures in mathematics and physics} \normalfont ed. C. DeWitt and J.A. Wheeler (Benjamin, New York 1968).

\bibitem{Kuchar80}            K.V. Kucha\v{r}, in {\it Quantum Gravity 2: A Second Oxford Symposium} ed. C.J. Isham, R. Penrose and D.W. Sciama (Clarendon, Oxford 1981).

\bibitem{Isham93}             C.J. Isham, in {\it Integrable systems, quantum groups and quantum field theories}  
                              ed. L.A. Ibort and M.A. Rodr\'{\i}guez (Kluwer, Dordrecht 1993), gr-qc/9210011.

\bibitem{KucharPOTother}      K.V. Kucha\v{r}, in {\it Conceptual Problems of Quantum Gravity} ed. 
                              A. Ashtekar and J. Stachel (Birkh\"{a}user, Boston, 1991);
                              K.V. Kucha\v{r}, in \it The Arguments of Time \normalfont
                              ed. J. Butterfield (Oxford University Press, Oxford 1999).

\bibitem{bigcite}             C.J. Isham, in Lect. Notes Phys. {\bf 434} (1994)  ,gr-qc/9310031; 
                              writeup of GR14 plenary lecture, gr-qc/9510063; 
                              C. Rovelli, Living Rev. Rel. {\bf 1} 1 (1998), gr-qc/9710008; 
                              J. Butterfield and  C.J Isham, in {\it Physics meets Philosophy at the Planck Scale}, ed. C. Callender and N. Huggett, Cambridge University Press (2000),  gr-qc/9903072;
                              S. Carlip, Rept. Prog. Phys. {\bf 64} 885 (2001), gr-qc/0108040;                                                           
                              T. Thiemann, Lect. Notes Phys. {\bf 631} 41 (2003), gr-qc/0210094; 
                              L. Smolin, hep-th/0303185;  hep-th/0408048;                              
                              A. Ashtekar and J. Lewandowski Class. Quant. Grav. {\bf 21} R53 (2004), gr-qc/0404018; 
                              O. Dreyer, hep-th/0409048.

\bibitem{Dirac}               P.A.M. Dirac, \it Lectures on Quantum Mechanics \normalfont (Yeshiva University, NY, 1964).

                                       
\bibitem{BSW}                 R.F. Baierlain, D. Sharp and J.A. Wheeler, Phys. Rev. \bf 126 \normalfont 1864 (1962).  

\bibitem{Vanderson}           E. Anderson, Phys. Rev. {\bf D68} 104001 (2003),  gr-qc/0302035.   
                             
\bibitem{TP}                  E. Anderson, ``Geometrodynamics: spacetime or space?"  (Ph.D. Thesis, University of London 2004), gr-qc/0409123;   
                              E. Anderson, gr-qc/0511070.  

\bibitem{SIPP}                J.B. Barbour, Class. Quantum Grav. \textbf{20}, 1543 (2003), gr-qc/0211021.

\bibitem{ABFKO}               E. Anderson, J.B. Barbour, B.Z. Foster, B. Kelleher and N \'{O} Murchadha, Class. Quantum Grav. {\bf 22} 1795 (2005),  gr-qc/0407104. 

\bibitem{LJR}                 See e.g. R.G. Littlejohn and M. Reinsch, Rev. Mod. Phys. {\bf 69} 213 (1997).  

\bibitem{Lanczos}             C. Lanczos, \it The Variational Principles of Mechanics \normalfont (University of Toronto Press, Toronto 1949; Dover, New York 1986).

\bibitem{Stewart}             J. Stewart, {\it Advanced General Relativity} (Cambridge University Press, Cambridge, 1991). 

\bibitem{BB77}                J.B. Barbour and B. Bertotti, Nuovo Cim. \bf B38 \normalfont 1 (1977).

\bibitem{HughesDrever}        V.W. Hughes, H.G. Robinson and V. Beltran--Lopez, Phys. Rev. Lett. {\bf 4} 342 (1960);                         
                              R.W.P. Drever, Phil. Mag. {\bf 6} 683 (1961).  

\bibitem{Marchal}             See e.g. C. Marchal, {\it Celestial Mechanics} (Elsevier, Tokyo 1990).

\bibitem{Xia}                 This was conjectured by P. Painlev\'{e}, {\it Le\c{c}ons sur la th\'{e}orie analytique des \'{e}quations diff\'{e}rentielles} (Hermann, Paris 1897);  
                                                            one example which has now been built is Z. Xia, Ann. Math {\bf 135} 411 (1992);
                              I mention this one in particular as it is for $\L = 0$; for other examples and a simple introduction see  
                              F.N. Diacu, {\it Singularities of the $N$-body problem} (Les Publications CRM, Montr\'{e}al, 1992).  

\bibitem{HillMoeckel}         See e.g. R. Moeckel, Contemp. Math. {\bf 81} 1 (1988) and references therein. 

\bibitem{v2}                  E. Anderson, v1 of the preprint gr-qc/0511068.  

\bibitem{MF}                  P.M. Morse and H. Feshbach, {\it Methods of Theoretical Physics Pert I} (McGraw--Hill, New York 1953).\bibitem{Robinett}            R.W. Robinett, {\it Quantum mechanics: classical results, modern systems, and visualized examples} 
                              (Oxford University Press, New York 1997). 

\bibitem{Landau}              L.D. Landau and E.M. Lifschitz, {\it Quantum Mechanics. Non-relativistic theory} 
                              (English translation: Pergamon, London, 1958).  

\bibitem{Schiff}              L.I. Schiff, {\it Quantum mechanics} (McGraw-Hill, New York 1968).  

\bibitem{Soland}              E. Anderson, {\bf 24} 2935 (2007), gr-qc/0611007;
                  E. Anderson, {\bf 24} 2971 (2007), gr-qc/0611008; 
                  E. Anderson 2 further forthcoming papers.  

\bibitem{Schwinger}           J. Schwinger {\it Quantum Mechanics} ed. B.-G. Englert (Springer, Berlin 2001).  

\bibitem{Mott}                N.F. Mott, {\it Elements of wave mechanics} (Cambridge University Press, Cambridge 1952); 
                              N.F. Mott and H.S.W. Massey, {\it The theory of atomic collisions} 
                              (Clarendon Press, Oxford 1949).

\bibitem{DiracQM}             See e.g. P.A.M. Dirac, {\it Principles of Quantum Mechanics} 4th edition 
                              (Clarendon Press, Oxford 1983).

\end{thebibliography}
\end{document}